\documentclass[review,authoryear]{elsarticle}
\usepackage{lineno,hyperref}
\modulolinenumbers[5]

\journal{Journal of \LaTeX\ Templates}

\newcommand\jgr{{J.~Geophys.~Res.}}%
\newcommand\ssr{{Space~Sci.~Rev.}}%
\newcommand\icarus{{Icarus}}%
\usepackage{subfigure}
\usepackage{here}
\biboptions{authoryear}

\begin{document}

\begin{frontmatter}

\title{Crystallization and cooling conditions for diogenite formation in the turbulent magma ocean of asteroid 4 Vesta}

%% Group authors per affiliation:
\author[mymainaddress,mysecondaryaddress]{Yusuke Kawabata}
\author[mymainaddress]{Hiroko Nagahara}

%% or include affiliations in footnotes:

\address[mymainaddress]{Department of Earth and Planetary Science, The University of Tokyo, 7-3-1 Hongo, Bunkyo-ku, Tokyo 113-0033, Japan}
\address[mysecondaryaddress]{Institute of Space and Astronautical Science, Japan Aerospace Exploration Agency, 3-1-1 Yoshinodai, Chuo, Sagamihara, Kanagawa 252-5210, Japan}

\begin{abstract}
Despite its small size, the asteroid 4 Vesta has been completely differentiated to core and mantle. Its composition is similar to howardite--eucrite--diogenite (HED) meteorites of which the detailed petrology is known. Therefore, Vesta is a good target for understanding the differentiation of terrestrial planets. A new differentiation model for crust formation has been developed by taking magma ocean fluid dynamics, chemical equilibrium, the presence of $^{26}$Al, and cooling into consideration with a special focus on crystal separation. The role of crystal size, thickness of the conductive lid, and fO$_{2}$ are evaluated as parameters. The results show that large crystals of at least 1 cm settled and formed a kilometer-thick cumulate layer of orthopyroxene with Mg$\#$ of 0.70-0.90 in $\sim$20 thousand years, which almost agrees with the Mg$\#$ of diogenites. Smaller grain sizes formed thinner layers. 
\end{abstract}


\end{frontmatter}

%\linenumbers

\section{Introduction}
Silicates and Fe-metal making up terrestrial planets were melted one or more times during the accretionary stage before reaching the final size. Giant collisions with bodies the size of Mars or the Moon could have melted and even partially vaporized the Earth, which has been thought to be the most plausible process for the origin of the Moon \citep{1986Sci...231..341B, 1987AREPS..15..271S}. Giant impacts and subsequent magma oceans caused a greenhouse effect in the atmosphere and core formation in the interior because volatile elements were contained \citep{1985LPSC...15..545A, 2008E&PSL.271..181E}. Magma oceans further caused compositional differentiation into mantle and crust, affected the concentration and distribution of volatile element, and controlled the thermal history of the planetary bodies \citep{2012AREPS..40..113E}.

The solidification processes of magma oceans determine the initial compositional differentiation of the silicate portions of planets, whereas the processes of crystallization and differentiation in magma oceans of terrestrial planets are described with fluid dynamics  \citep{1993LPI....24.1329S, 1993JGR....98.5407S, 1993JGR....98.5375S}. It has been suggested that a low-viscosity, vigorously convecting magma
ocean is more similar to the atmosphere rather than to the solid mantle.

Whether growing mineral grains remain suspended in the melt or settle out is fundamental to the chemical evolution of a magma ocean. The process of solidification can be viewed simply as two end members, namely fractional solidification and batch solidification, with natural processes occurring between them. Fractional solidification in a magma ocean requires separation of mineral grains from the flow and effective isolation from interaction with the remaining magma ocean liquids. However, batch solidification maintains the interaction of liquid and crystallized minerals throughout the solidification process.

Crystal settling is a fundamental process in a magma ocean of
a small planetary body. Crystallization occurs in all regions owing to
the almost parallel temperature relationship between adiabat and liquidus, where the density contrast between minerals and liquid is of primary importance. Although the densities of minerals and liquid depend on chemical composition, pressure, and temperature, the dependence differs between solid and liquid, which needs a chemical equilibrium calculation consistent with the fluid dynamics of the magma ocean.

The fluid dynamics of a turbulent magma ocean include crystal separation, formation of a mush layer, and entrainment of those minerals into liquid. All of these processes are strongly affected by the size, shape, and mutual relationship of multiple minerals and are controlled by cooling of the magma ocean \citep{2000orem.book..323S, solomatov2007magma}.

Asteroid 4 Vesta is the only preserved intact example of a large,
differentiated protoplanet \citep{2012Sci...336..684R}. It has a basaltic crust, and observations with reflectance spectra have provided convincing evidence for a differentiated interior that includes an ultramafic mantle exposed in the cavity of a massive impact basin and a possible iron-rich core \citep{2012Sci...338..242P}. The reflectance spectra of Vesta  \citep{2013M&PS...48.2166D} show
many similarities to those of howardite--eucrite--diogenite (HED) meteorites, 
where howardite is a breccia of eucrite and diogenite, eucrite is a
basaltic achondorite, and diogenite is an orthopyroxenite. This suggests that Vesta
is the parent body of HED meteorites. Geochemical, petrological, and geochronological studies have revealed that Vesta was melted substantially possibly by the decay of $^{26}$Al and $^{60}$Fe  \citep{2011M&PS...46..903M,2012espc.conf..909N, 2014E&PSL.395..267N}. Crystallization modeling of the magma ocean of Vesta has produced
an olivine mantle, a lower crust rich in diogenites, and an upper crust of
basaltic flows and eucrites \citep{1997M&PS...32..929R}. Recently, Dawn
spacecraft observed the surface mineralogy of Vesta with the visible and infrared
spectrometers (VIRs), the latter of which revealed spatially resolved
hyperspectral images of Vesta \citep{2011SSRv..163..329D}. The results enabled us to estimate regionally localized
mineralogical units suggestive of a complex geological and collisional history.

Whether Vesta was once melted entirely or partially remains controversial. \cite{1997M&PS...32..929R} and \cite{2013M&PS...48.2333M} proposed a model of the entirely melted magma ocean and discussed whether the mode of crystallization was equilibrium or fractional. On the contrary, \cite{2012espc.conf..909N, 2014E&PSL.395..267N} discussed the difficulty of total melting of small bodies and suggested a model of partially melted mantle including the transport of $^{26}$Al, a critical heat source, to the upper portion.

In the present work, we investigate the evolution of the magma ocean at the final stage, and we focus on the roles of physical parameters affecting the evolution of the magma ocean, grain size of crystals, thickness of the lid, and fO$_{2}$. The grain size of crystals is related to the rate of crystal settling, which controls the rate of chemical differentiation. The thickness of the lid directly controls the cooling time scale of the magma ocean, and the composition of the minerals and melt varies with fO$_{2}$. In this study, we evaluate the roles of the three parameters on solidification of a magma ocean under a turbulent flow by consistently calculating the heat balance, settling of crystals, and chemical compositions of melt and crystals.

The rest of this paper is organized as follows. A description of our model is given in Section 2, the results of the numerical simulation are shown in Section 3, and the results are discussed in Section 4. Finally, we summarize our study in Section 5.

\section{Model setup}
In order to investigate the final evolution of the magma ocean of Vesta, we developed a numerical model combining fluid dynamics and thermodynamics. A plausible range of parameters is constrained, which indicates an orthopyroxenite layer of considerable thickness in which the chemical composition is consistent with that of diogenites.
\subsection{Configuration of the model} 
 
In this study, we assumed that the shape of Vesta is spherical and that its interior had already been differentiated to form a core. The thickness of the core, the mantle, and the crust is often assumed to be 120 km, 80 km, and 50 km, respectively \citep{2011SSRv..163...77Z}. After the core differentiated, the mantle of Vesta was totally molten. With subsequent cooling, olivine crystallized and settled down to form the mantle. The melt of the partially molten mantle moved to the surface and formed a shallow magma ocean.   
The timing of the olivine extraction remains controversial. \cite{1997M&PS...32..929R} reported that the olivine extraction occurred after 80\% equilibrium crystallization. However, \cite{2013M&PS...48.2333M} estimated 60--70\% on the basis of \cite{1962PhFl....5.1374K}. The magma ocean would, however, have been viscous enough to suppress the velocity of the convection before the regime. Therefore, we assumed that the olivine extraction occurred after 55\% equilibrium crystallization, and the remaining melt fraction, 45\%, corresponds to the mass of the crust of Vesta. The critical value was experimentally determined by \cite{van1979experimental}, and \cite{lejeune1995rheology}, in which 55\% is the most typical value. We set the initial value of the thickness of the shallow magma ocean as 50 km, which is close to that of the crust of Vesta obtained by observation. The calculation of the present work began after the olivine extraction and the formation of the shallow magma ocean.  

\begin{figure*}
\centering
 \includegraphics[width=\columnwidth,clip,bb=0.000000 0.000000 717.000000 616.000000]{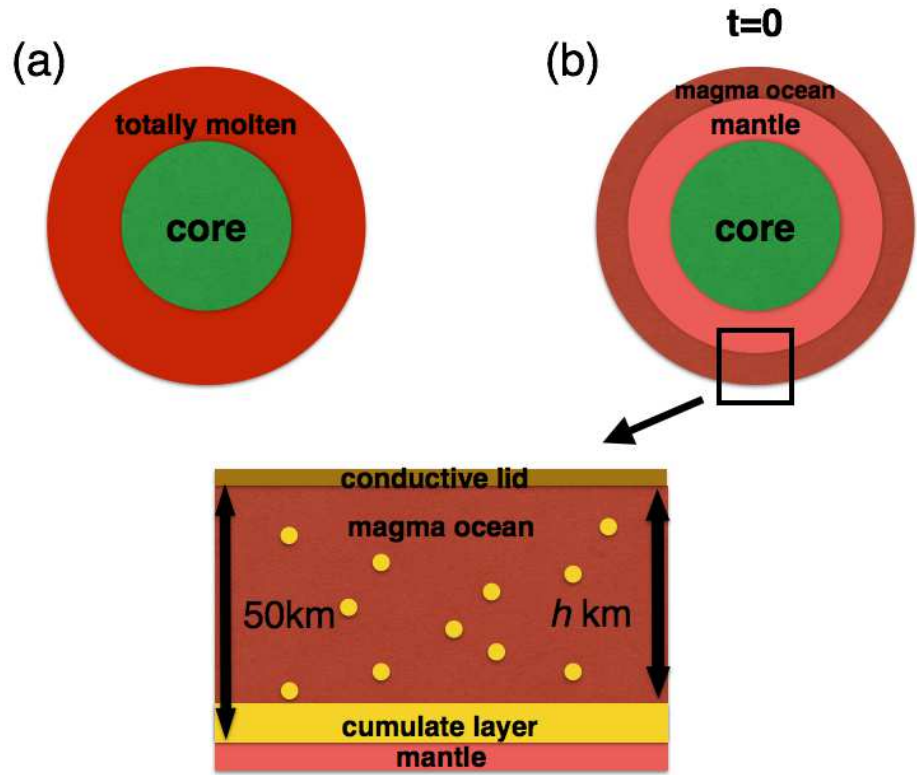}
 \caption{ Model configuration of the magma ocean of Vesta in the present work.}
 \label{model}
 \begin{flushleft}
 (a) After the core differentiated, the silicate portion was totally molten. (b) With subsequent cooling, olivine crystallized and settled down to form the mantle. The calculation of the present work starts from this point. The melt of partially molten mantle moved to the surface and formed the shallow magma ocean. We assumed the initial depth of the magma ocean to be 50 km. The depth of the magma ocean became thinner as the crystals settled. During solidification, the distribution of the particles was uniform because of the turbulence.
 \end{flushleft}
\end{figure*}

The schematic illustration of the model is shown in Figure \ref{model}. Furthermore, we considered the existence of a lid above the shallow magma ocean. Although \cite{2013M&PS...48.2333M} determined the thickness of the lid to be 250--750 m, this estimation remains controversial. Therefore, we varied the thickness of the lid as a parameter to be 1 km, 100 m, and 10 m. If the conductive lid is thick and evolves with time, the lid could have an influence on the chemistry of the magma ocean. In our study, the boundary layer was assumed to be quenched melt; hence, the conductive lid did not influence the chemistry of the magma ocean because the lid and melt have the same composition.

All of the parameters used in this model are summarized in Table \ref{tb:parameter}. The mass of Vesta was determined from its perturbation by other asteroids, and the shape was derived from Hubble Space Telescope images \citep{1997Sci...277.1492T}. The physical parameters of Vesta summarized by  \cite{2011SSRv..163...77Z} were used in the present study. Other physical parameters, thermal expansion, thermal capacity, $^{26}$Al half-life, and latent heat of the magma ocean were taken from \cite{2000orem.book..323S, solomatov2007magma} and \cite{2012espc.conf..909N, 2014E&PSL.395..267N}.

\subsection{Heat transfer}
A convective layer of a fluid is generally described with dimensionless parameters, a Rayleigh number ($Ra$), and a Prandtl number ($Pr$), defined as
\begin{eqnarray}
Ra&=&\frac{\alpha g(T-T_{{\rm b}} ) h^3}{\kappa\nu}     \\
Pr&=&\frac{\nu}{\kappa},
\end{eqnarray}
where $\alpha$ is the thermal expansion and $T$ is the temperature of the shallow magma ocean. The temperature in the entire magma ocean can be regarded as being uniform during the cooling because the magma ocean is highly convective where the temperature profile is along the adiabatic profile and because the size of the planetary body is small enough for which the effect of pressure can be neglected. $T_{{\rm b}}$ is the temperature of the top of the shallow magma ocean, i.e., the bottom of the quenched lid.  $h$ is the depth of the fluid layer, $\kappa$ is the thermal diffusivity, and $\nu$ is the kinematic viscosity. The Rayleigh number characterizes the vigor of the convective flow, which is the ratio of the buoyancy to the viscosity and thermal diffusion. The convection becomes vigorous with an increase in the Rayleigh number. The Prandtl number is the ratio of kinematic viscosity to the thermal diffusivity and measures the effectiveness of thermal diffusion in regulating the flow.

Convection has two regimes: soft turbulence and hard turbulence \citep{2000orem.book..323S, solomatov2007magma}. Their heat fluxes are described as
\begin{eqnarray}
F_{\rm{soft}}&=&0.089 \frac{k(T-T_{{\rm b }})}{h} Ra^{1/3} \label{eq:soft} \\
F_{\rm{hard}}&=&3.43\times10^{-3}  \frac{k(T-T_{{\rm b}} )}{h} Ra^{3/7} Pr^{-1/7} \lambda^{-3/7}, \label{eq:hard}
\end{eqnarray}
where $\lambda$ is the aspect ratio for the mean flow \citep{1962PhFl....5.1374K, 1994AnRFM..26..137S}. These modes of convection change according to the magnitude of the Rayleigh number.
We determined whether the turbulence is soft or hard according to the condition reported by \cite{so24778}:
\begin{equation}
4.8\times10^{-8}Ra^{2/3}\geq Pr . \label{eq:condition}
\end{equation}

When condition (\ref{eq:condition}) is satisfied, the mode of convection changes from soft to hard. This change in the convection mode affects the rate of cooling. In our model, the Rayleigh number is $\sim10^{18}$ at the beginning and decreases with cooling. When the crystal fraction of the shallow magma ocean reaches $\sim$10\%, depending on the parameters, the relationship between the Rayleigh number and the Prandtl number no longer satisfies the condition (\ref{eq:condition}), and the mode of the convection changes to soft.

The heat flux of the convection must match the conductive heat flux propagating the lid, which is expressed as Fourier's law:
\begin{equation}
F_{\rm{cond}}=-k\nabla T=-k\frac{\Delta T}{l}, \label{eq:conduction}
\end{equation}
where $l$ is the thickness of the lid above the magma ocean, $\Delta T= T_{{\rm s}}- T_{{\rm b}}$, and $T_{{\rm s}}$ is the temperature of the upper part of the quenched crust.
The heat flux of the conduction must match the surface heat flux, which can be calculated with the help of the
blackbody radiation:
\begin{equation}
F_{\rm{rad}}=\sigma T_{{\rm s}}^4.\label{eq:stefan}
\end{equation}
From Eqs. (\ref{eq:soft}), (\ref{eq:hard}), (\ref{eq:conduction}), and (\ref{eq:stefan}), we solve the temporal evolution of the temperature of the magma ocean. The initial temperature of the magma ocean T is defined by the bottom and the surface temperatures. The bottom temperature is defined by the appearance of pyroxene when the magma ocean depth decreases to 50 km, which corresponds to 1670 K. We calculated the $T_{b}$ and $T_{s}$ computed by equating the blackbody radiation with the conductive heat flux and convective flux. We fixed the time step at 100 years and solved the four equations so that each flux was equivalent at each boundary by Newton's method. As previously mentioned, the temperature of the magma ocean was assumed to be uniform. That is, we calculated the heat flux only at each boundary. When solving the equations, we also considered the heat flux by the radio decay of ${}^{26}$Al (Neumann et al., 2012 ) and the latent heat of the crystallization:
\begin{eqnarray}
&F_{{\rm Al}}&=M_{{\rm magma}}f_{{\rm Al}}\left[\frac{^{26}Al}{^{27}{Al}}\right]\frac{E_{Al}}{\tau_{1/2}}e^{\frac{-t+t_{0}}{\tau_{1/2}}} \\
&F_{{\rm latent}}&=Q_{{\rm latent}}\Delta M_{{\rm crystal}},
\end{eqnarray}
where $M_{{\rm magma}}$ is the mass of the magma ocean, $f_{{\rm Al}}$ is the abundance of Al, $^{26}$Al/$^{27}$Al is the initial abundance ratio, $\tau_{1/2}$ is the half-life,  $t_{0}$ is the time after the formation of Ca--Al-rich inclusions, $Q_{{\rm latent}}$ is the latent heat per unit mass of the crystals, and $\Delta M_{\rm crystal}$ the mass of the crystals at a  certain time step. Equations (8) and (9) were used for calculating the temporal change of the temperature as an input energy to the magma ocean. The distribution of $^{26}$Al was assumed to be the same as that of Al. The distribution of Al is described by calculation of the MELTS program, as shown below. We neglected the heat flux from the core and the mantle in our model.

\subsection{Crystallization and crystal settling}
We calculated the phase relations from the liquidus to solidus temperature by using the MELTS program \citep{ghiorso1995chemical, asimow1998algorithmic}. MELTS, a software package designed to facilitate the thermodynamic modeling of phase equilibria in natural magmatic systems, was constructed by optimizing nearly 2500 solid--liquid experiments.

We assumed that the composition of the magma ocean of Vesta after core formation is a silicate portion of L-chondrite after \cite{1997M&PS...32..929R}, which is shown in the first column of Table \ref{tb:composition}. The initial composition for the present model calculation was obtained by crystallization and extraction of olivine at fO$_{2}$ $=$ quartz--fayalite--magnetite (QFM) $\pm 2$, QFM $\pm 1$, and iron--wustite (IW) by MELTS until the depth of the magma ocean decreased to 50 km, which are shown in Table \ref{tb:composition}. The magma ocean was regarded to be uniform in terms of chemical composition because it is vigorously convective.

The crystal separation rate from the turbulently convective fluid, as discussed by \cite{martin1989fluid}, assumes the following:
(a)the crystals are distributed homogeneously in the flow;
(b)there is no re-entrainment of crystals into the flow once they have settled out;
(c)crystals influence neither each other nor the nature of the melt; and
(d)the grain size is uniform. 
We applied their model to our crystal separation model, as shown in Eqs. (10)--(16).

The convective velocity vanishes at the bottom of the melt, where the crystal
movement stops. The rate of decrease in the number of crystals in the flow with time is given by
\begin{eqnarray}
\frac{dN}{dt}&=& -A v_{\rm{s}} c(0) \label {eqn:martin} \\
v_{s} &=&\frac{g\Delta \rho a^{2}}{18\rho \nu}, \label{eqn:stokes}
\end{eqnarray}
where $N$ is the crystal number, $A$ is the area of the bottom of the magma ocean, $v_{\rm{s}}$ is the crystal settling
velocity given by Stokes' Law, $c(z)$ is the concentration of crystals at height $z$ above the bottom boundary, $g$ is the acceleration owing to gravity, $a$ is the diameter of the crystal, $\Delta\rho$ is
the density contrast between the crystal and the melt, and $\nu$ is the
kinematic viscosity of the melt.
Because the crystals are uniformly distributed in the most parts of the flow,
The following can be assumed:
\begin{equation}
c(0) \sim \frac{N}{Ah}, \label{eqn:bottom}
\end{equation}
where h is the depth of the magma ocean. Substitution of Eqs. (\ref{eqn:stokes}) and (\ref{eqn:bottom}) into (\ref{eqn:martin})
gives
\begin{equation}
\frac{dN}{dt}=N\left(\frac{-g\Delta \rho a^{2}}{18\rho \nu h} \right). \label{eqn:martin2}
\end{equation}

Assuming that the grains are spherical, the particle number N satisfies
\begin{equation}
N=\frac{6\phi V}{\pi a^{3}},
\end{equation}
where $\phi$ is the crystal fraction, and $V$ is the volume of the magma ocean. The thickness of the magma ocean $h$ decreases with time by the settlement of the particles. The
decrease in the depth of the melt layer $\Delta h$ without interstitial melt is
\begin{equation}
\frac{4}{3}\pi(r_{\rm{c}} + \Delta h)^{3} = V_{\rm{c}} +\frac{\pi a^{3}}{6}N_{\rm {settle}}, \label{eqn:settling}
\end{equation}
where $N_{\rm{settle}}$ is the number of the crystals settled, and $r_{\rm{c}}$ and $V_{\rm{c}}$ are the radius and the volume of the core and the unmelted layer, respectively. Rewriting Eq. (\ref{eqn:settling}), we obtain
\begin{equation}
\Delta h = (\frac{3}{4}\pi V_{\rm{c}} +\frac{a^{3}}{8}N_{\rm{settle}})^{1/3}-r_{\rm{c}}.
\end{equation}

For clarity, the calculation scheme of our study is summarized as follows:
\begin{description}
\item[(i)] Calculate the heat flux $F$ using Eqs. (\ref{eq:soft}), (\ref{eq:hard}), (\ref{eq:conduction}), and (\ref{eq:stefan}) at a certain time step.
\item[(ii)] By using $F$, find the temporal change in temperature during the time step.
\item[(iii)] Calculate the number of settling crystals $N_{{\rm settle}}$ and the decrease of the depth of the magma ocean $\Delta h$,
\item[(iv)] Input the new composition (previous composition minus settled particle) to MELTS, and then return to (i).
\end{description}
We performed these calculations by varying the value of the diameter of the crystal, the thickness of the conductive lid, and ${\rm fO_{2}}$ as parameters.
\section{Results}
\subsection{Cooling of the magma ocean and cumulate layer formation}

\begin{figure*}
 \centering
   \subfigure[]{
  \includegraphics[width=0.5\columnwidth,clip,bb=0.000000 0.000000 850.009733 850.009733]{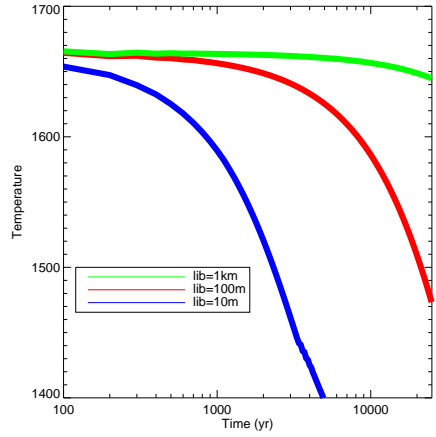}
  }
  \subfigure[]{
 \includegraphics[width=0.5\columnwidth,clip,bb=0.000000 0.000000 850.009733 850.009733]{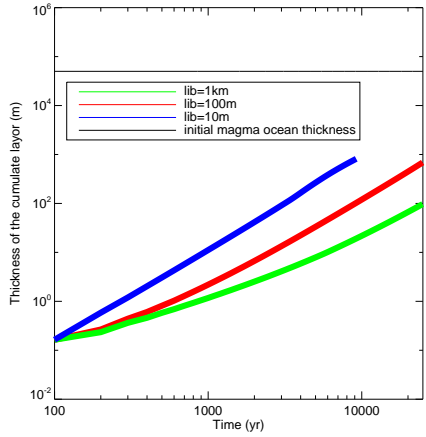}
 }\label{lida}

 \caption{(a)Temporal change in the temperature of the magma ocean with the thickness of the lid as a parameter. The lid at 1 km, 100 m, and 10 m are displayed by green, red, and blue lines, respectively.(b)Temporal change in the thickness of the cumulate layer with lid thickness as a parameter. The lids at 1 km, 100 m, and 10 m are displayed in the same colors as those in (a). The diameter of the crystal is 0.1 cm and ${\rm fO_{2}}$ is IW for both (a) and (b).}
 \small{}
 \label{lid}
\end{figure*}

\begin{figure*}
 \centering
 \subfigure[]{
  \includegraphics[width=0.5\columnwidth,clip,bb=0.000000 0.000000 850.009733 850.009733]{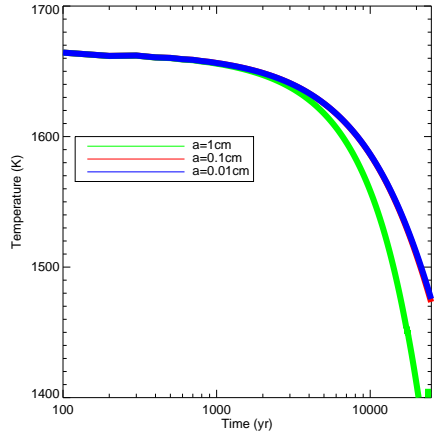}}
   \subfigure[]{
 \includegraphics[width=0.5\columnwidth,clip,bb=0.000000 0.000000 850.009733 850.009733]{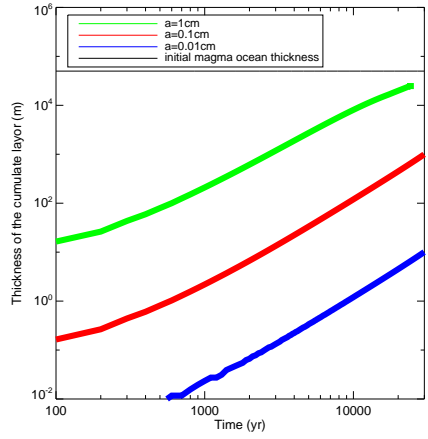}}
 \caption{(a)Temporal change in temperature of the magma ocean; (b)temporal change in thickness of the magma ocean with grain size diameter as a parameter. Diameters 0.01 cm, 0.1 cm, and 1 cm are represented by blue, red, and green lines, respectively. The thickness of the lid is 100 m, and ${\rm fO_{2}}$ is IW for both (a) and (b).}
 \small{}
 \label{particle}
\end{figure*}

The temporal change in temperature of the magma ocean and the thickness of the cumulate layer are shown in Figs. \ref{lid} and \ref{particle}. It should be noted that our calculation began at the time of orthopyroxene appearance in the residual magma ocean after mantle olivine separation.

Figure \ref{lid} shows the change in magma ocean temperature (a) and the growth of the cumulate layer (b) where the crystal size is fixed at 0.1 cm and fO$_{2}$ at IW with the thickness of the conductive lid being a parameter from 10 m to 1 km. The crystalline phase for forming the cumulate layer is orthopyroxene, which will be subsequently discussed in detail. The rate of temperature decrease is strongly dependent on the thickness of the lid, as shown in Fig. \ref{lid}a. A lid 10 m thick cools to near the solidus temperatures within a few tens of thousands of years, although the cooling is much slower with a lid 1 km thick. Rapid temperature decrease with a thin lid is no surprise because the conductive heat flux increases as the lid becomes thinner. The thickness of the cumulate layer increases as the conductive lid becomes thinner, which is directly related to the cooling of the magma ocean, as shown in Fig. \ref{lid}a. The magma ocean cools very slowly with a thick lid; consequently, the growth rate of the cumulate layer is small. However, this tendency is the case only for the early stage of magma ocean crystallization. As time passes, the magma ocean with a thicker lid makes a thicker cumulate layer. Figure \ref{lid} indicates that the thickening of a lid prolongs the evolution time of the magma ocean such as the growth time of a cumulate layer as far as the grain size of crystals are independent of the cooling time scale.

The role of the grain size of minerals is shown in Fig. \ref{particle}, for which the diameter of the crystals varied from 0.01 cm to 1 cm with a fixed lid thickness of 100 m and fO$_{2}$ at the IW buffer. Figure \ref{particle}a shows the cooling of the magma ocean, and Fig. \ref{particle}b shows the temporal change in the thickness of the cumulate layer. The magma ocean cools with similar rates for the three cases (Fig. \ref{particle}a); however, the thickness of the cumulate layer differs significantly. The cumulate layer thickness varies by two orders of magnitude with a difference in grain size by one order, which is attributed to the difference in the settling velocity (Eqs. \ref{eqn:stokes} and \ref{eqn:martin2}). Figure \ref{particle}b shows that the crystals with diameters of 0.01 cm (the blue line in Fig. \ref{particle}) hardly settle down to form a cumulate layer; that is, most of the crystals are suspended in the magma ocean. Figure \ref{particle}a shows that the magma ocean with crystal diameters of 1 cm cooled faster than those with smaller sizes, whereas those with diameters of 0.1 cm and 0.01 cm showed no difference in the cooling rate. With a diameter of 1 cm, hard turbulence changes to soft at 1537 K. If a large number of crystals is settled out, the magma ocean becomes thinner, resulting in an increase in the cooling rate. No difference in the cooling rate between diameters of 0.1 cm and 0.01 cm means that the crystal separation rate is very slow compared with the cooling rate. 

In summary, crystallization proceeds almost as a batch process, maintaining chemical equilibrium between crystals and melt if the size of the crystals is smaller than $\sim$1 mm. On the contrary, the magma ocean cools rapidly and chemical fractionation is effective if the crystal size is larger than $\sim$1 cm even if the lid is $\sim$100 m thick. Therefore, the size of crystals plays a crucial role in the evolution of the magma ocean.

\subsection{Petrology of magma ocean and diogenite formation}
\begin{figure*}
 \centering
 \subfigure[lid $l$ = 100 m, diameter $a$ = 0.0 1cm]{
 \includegraphics[width=0.45\columnwidth,clip,bb=0.000000 0.000000 501.051494 325.033404]{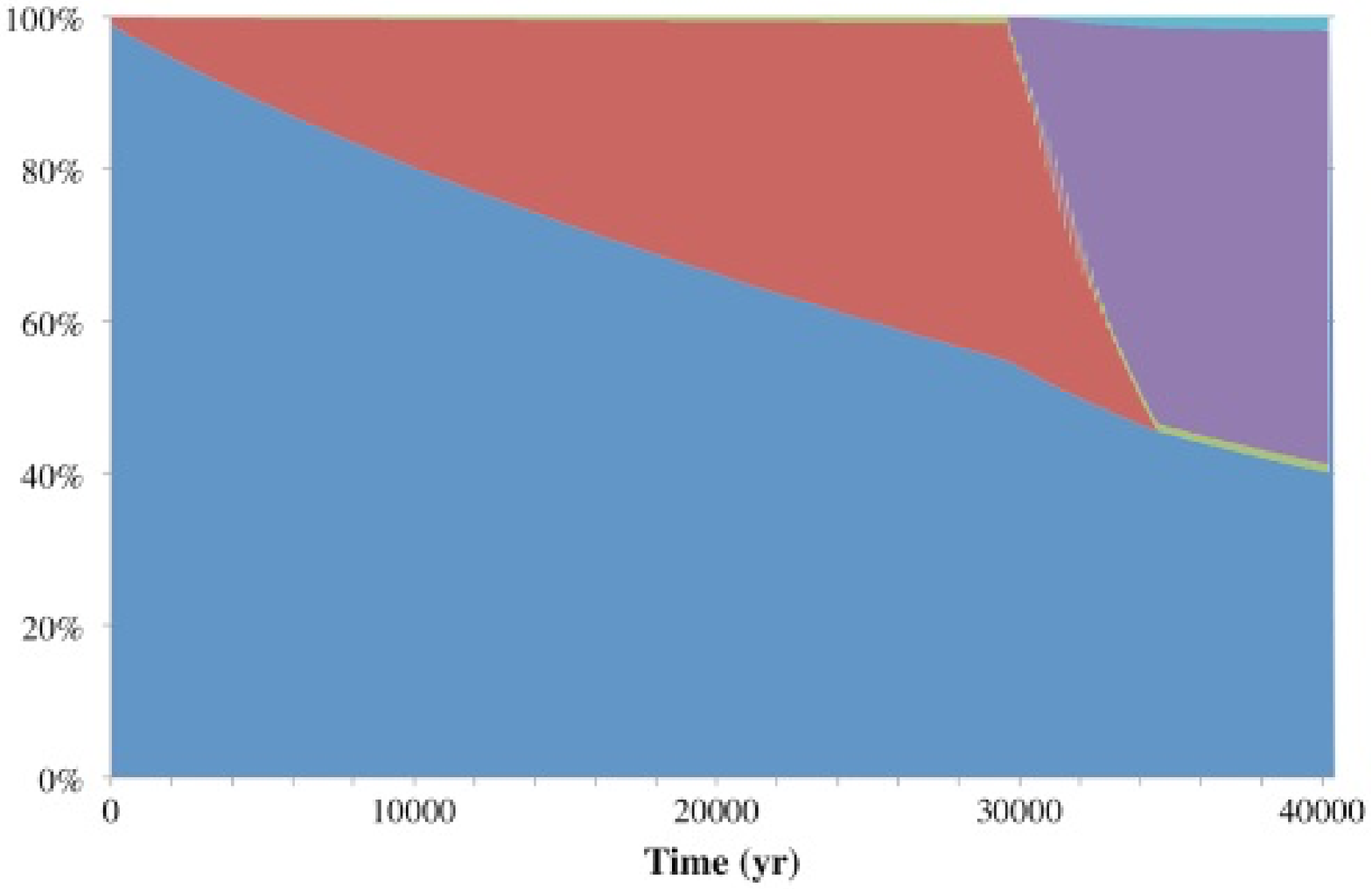}}
 \subfigure[lid $l$ = 100 m, diameter $a$ = 0.1 cm]{
 \includegraphics[width=0.45\columnwidth,clip,bb=0.000000 0.000000 501.051494 325.033404]{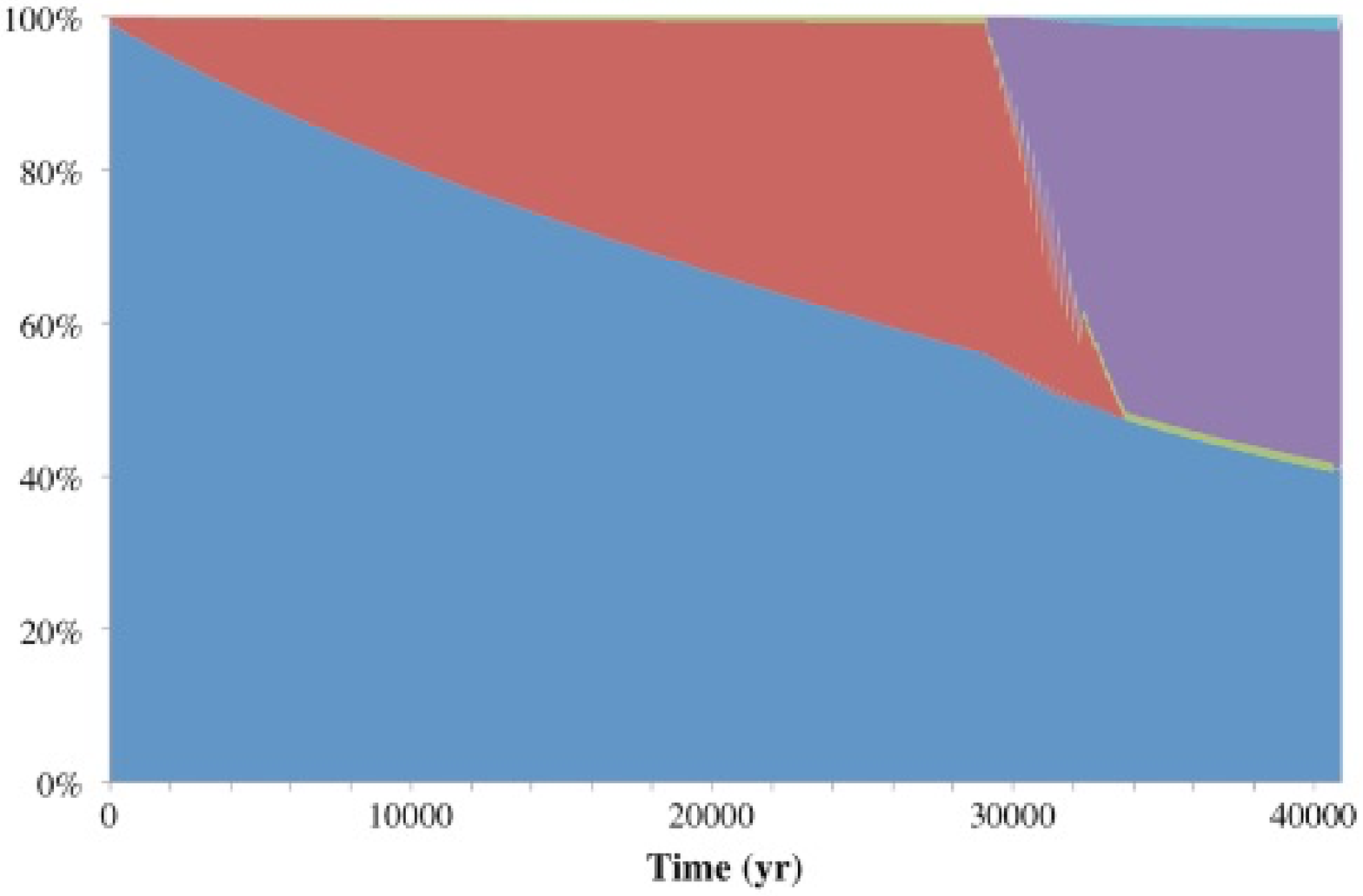}}
  \subfigure[lid $l$ = 100 m, diameter $a$ = 1 cm]{
 \includegraphics [width=0.45\columnwidth,clip,bb= 0.000000 0.000000 501.051494 325.033404]{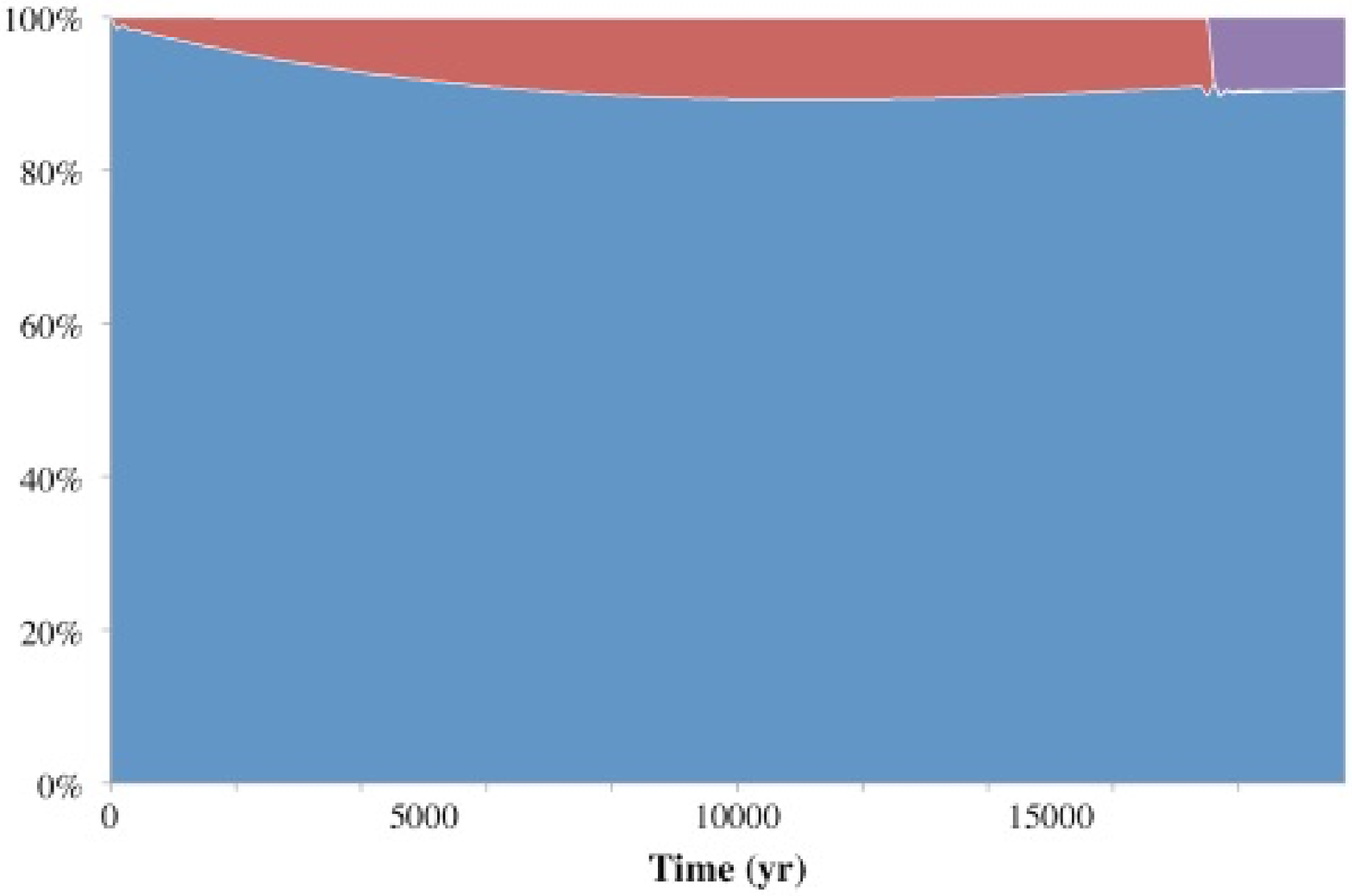}}
  \subfigure[lid $l$ = 1 km, diameter $a$ = 0.1 cm]{
 \includegraphics [width=0.45\columnwidth,clip,bb= 0.000000 0.000000 501.051494 325.033404]{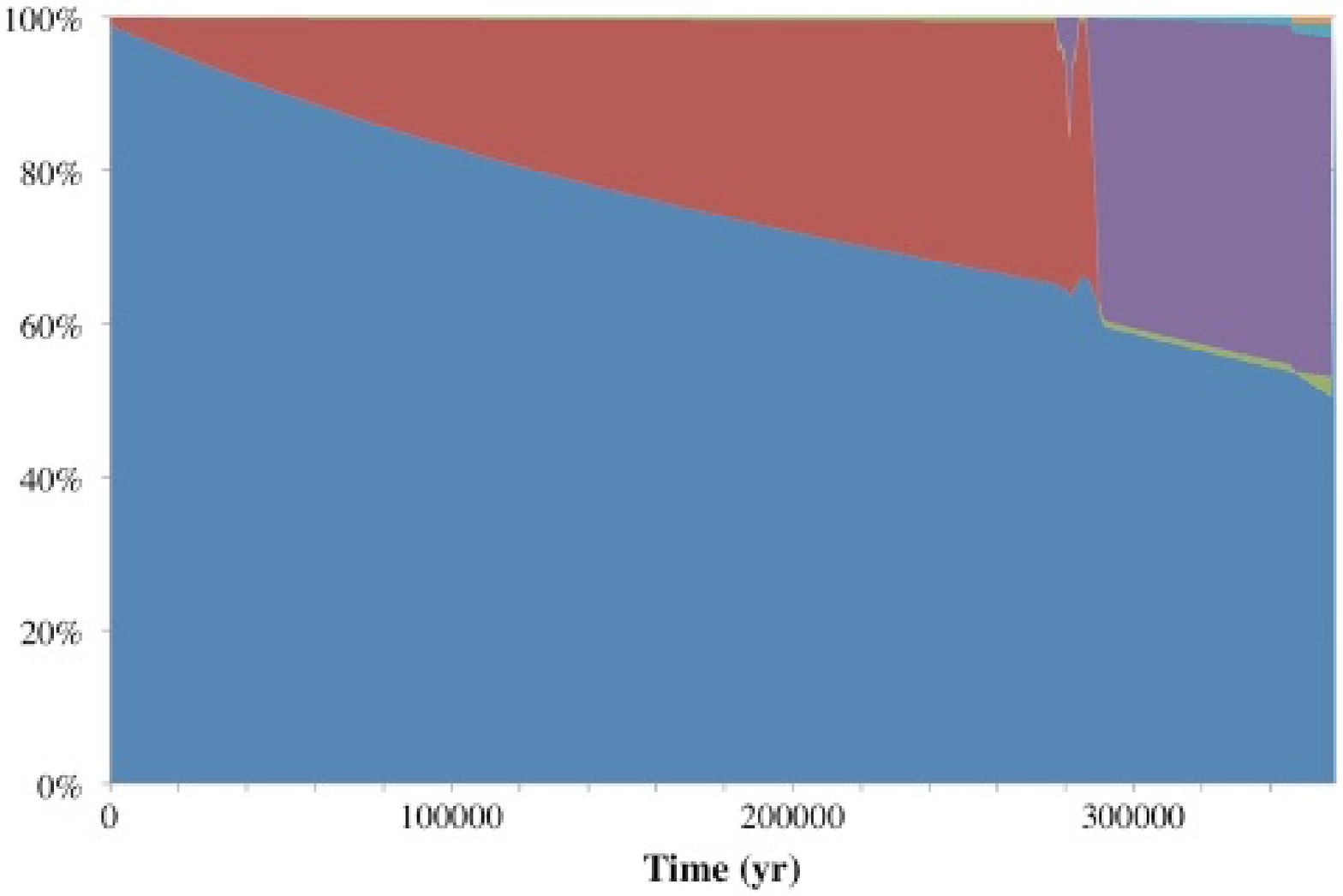}}
 \subfigure{
  \includegraphics[width=0.8\columnwidth,clip,bb=0.000000 0.000000 397.020402 21.001079]{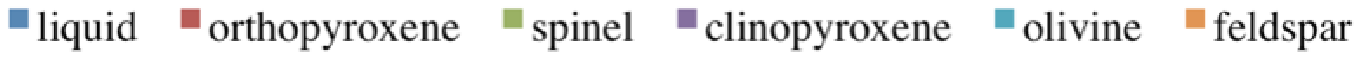}}
 \caption{Temporal change in the phase relations of the magma ocean; the crystal fractions are the same as those in the cumulate layer in the present model. The areas of blue, red, green, purple, light blue, and orange express the relative fraction of liquid, orthopyroxene, spinel, clinopyroxene, olivine, and feldspar, respectively. The parameters of each diagram are (a) lid $l$ = 100 m, diameter $a$ = 0.01 cm; (b) lid $l$ = 100 m, diameter $a$ = 0.1 cm; (c) lid $l$ = 100 m, diameter $a$ = 1 cm; (d) lid $l$ = 1 km, diameter $a$ = 0.1 cm.}
  \label{phase}
\end{figure*}

In this section, we show the chemical aspects of our calculation. As previously shown in Fig. \ref{particle}, the physical parameters of the magma ocean, specifically the grain size of crystals, largely affect the evolution of the magma ocean. This in turn affects the chemical compositions of the minerals and melts. Figure \ref{phase} shows the temporal phase change of the magma ocean after the appearance of orthopyroxene. It should be noted again that a significant amount of olivine has already crystallized before this point. Moreover, the fraction in the vertical scale represents the relative amount of minerals and melt in the uppermost 50 km at the beginning and less at the later stage, which is shown in Figs. \ref{lid} and \ref{particle}. Figure \ref{phase} shows the relative amounts of minerals and melt in the magma ocean alone, which does not include separated minerals. The physical parameters for each diagram are (a) lid $l$ = 100 m, diameter $a$ = 0.01 cm; (b) lid $l$ = 100 m, diameter $a$ = 0.1 cm; (c) lid $l$ = 100 m, diameter $a$ = 1 cm; and (d) lid $l$ = 1 km, diameter $a$ = 0.1 cm. Oxygen fugacity is QFM at 200 bar for all four panels. It should be noted that Fig. \ref{phase}a and \ref{phase}b are expressed in the same time scale, but Fig. \ref{phase}c and Fig. \ref{phase}d are at different time scales. Orthopyroxene (red) crystallizes first, followed mainly by clinopyroxene (purple) and much lesser amounts of spinel or plagioclase in many cases. Figures \ref{phase}a and \ref{phase}b, which have the same lid thickness with different crystal sizes, are similar, suggesting similar chemical evolution of the magma ocean. This is consistent with the observation that there is little difference in the cooling speed and the rate of crystal settling for the two cases, as shown in Fig. \ref{particle}b. However, Fig. \ref{phase}c is quite different from Fig. \ref{phase}a and \ref{phase}b, where the amount of orthopyroxene does not increase with time and is replaced by clinopyroxene after a considerable time. A comparison of Figs. \ref{particle}b and \ref{phase}c reveals that most of orthopyroxene crystallized settled down to the bottom of the magma ocean and that the residual magma ocean retained very small numbers of crystals, as shown in Fig. \ref{phase}c. The large melt area (blue) in Fig. \ref{phase}c does not represent the delay of crystallization, which cools rather rapidly as shown in Fig. \ref{particle}a. Figures \ref{phase}a, \ref{phase}b, and \ref{phase}c compare the role of grain size on the chemical evolution and reveal that the effective extraction of orthopyroxene for the case of large grain size (Fig. \ref{phase}c) put the appearance of clinopyroxene ahead. A thick lid did not affect the process significantly, with minor differences in the consumption of orthopyroxene and the appearance of clinopyroxene (straight boundary between red and purple in Fig. \ref{phase}d). 

From these results, we determined a strong dependence of the crystal settling and cooling rate on the crystal size. As time progressed, orthopyroxene crystallization stopped, and clinopyroxene began to appear. We will discuss the formation of orthopyroxenite and the dependence on parameters in the following section.
\section{Discussion}
In this section, we discuss the evolution of the magma ocean of 4 Vesta and compare our model with previous models.
\subsection{Orthopyroxenite layer and Mg$\#$}
The mineral phases and their abundance ratios in the cumulate layer are the same as those in the residual magma ocean in the present model because all of the minerals are assumed to have the same grain size, resulting in settling and maintaining their relative abundance ratios. Therefore, most of the cumulate layer is orthopyroxenite with small amounts of pyroxenite and clinopyroxenite. The sharp boundary between orthopyroxene and clinopyroxene in Figs. \ref{phase}c and \ref{phase}d suggests that pyroxenite is almost lacking in these cases. 

\begin{figure*}
 \centering
 \subfigure[diameter $a$=1cm]{
 \includegraphics[width=0.45\columnwidth,clip,bb=0.000000 0.000000  850.009733 850.009733]{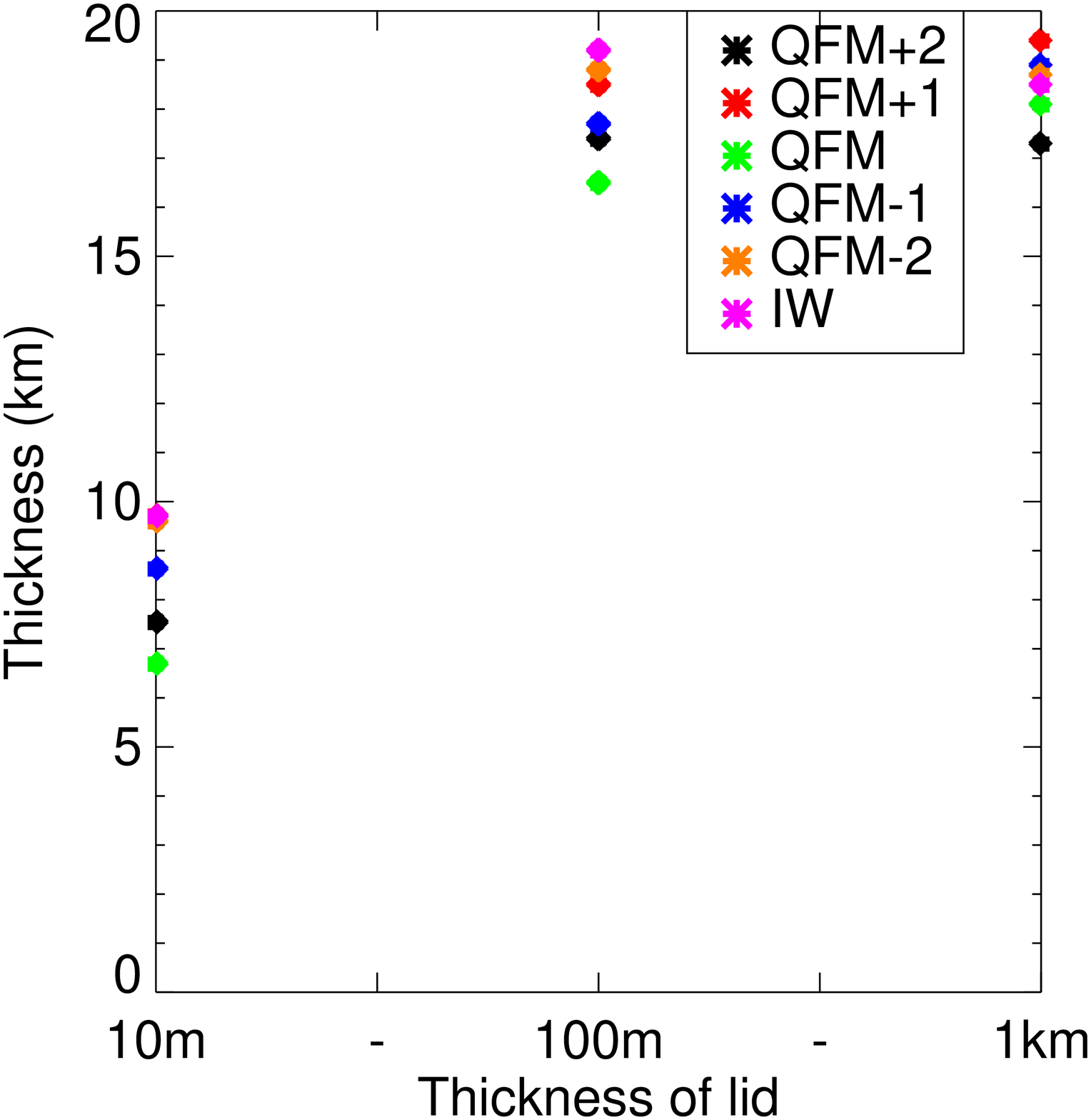}}
 \subfigure[diameter $a$=0.1cm]{
 \includegraphics[width=0.45\columnwidth,clip,bb=0.000000 0.000000 850.009733 850.009733]{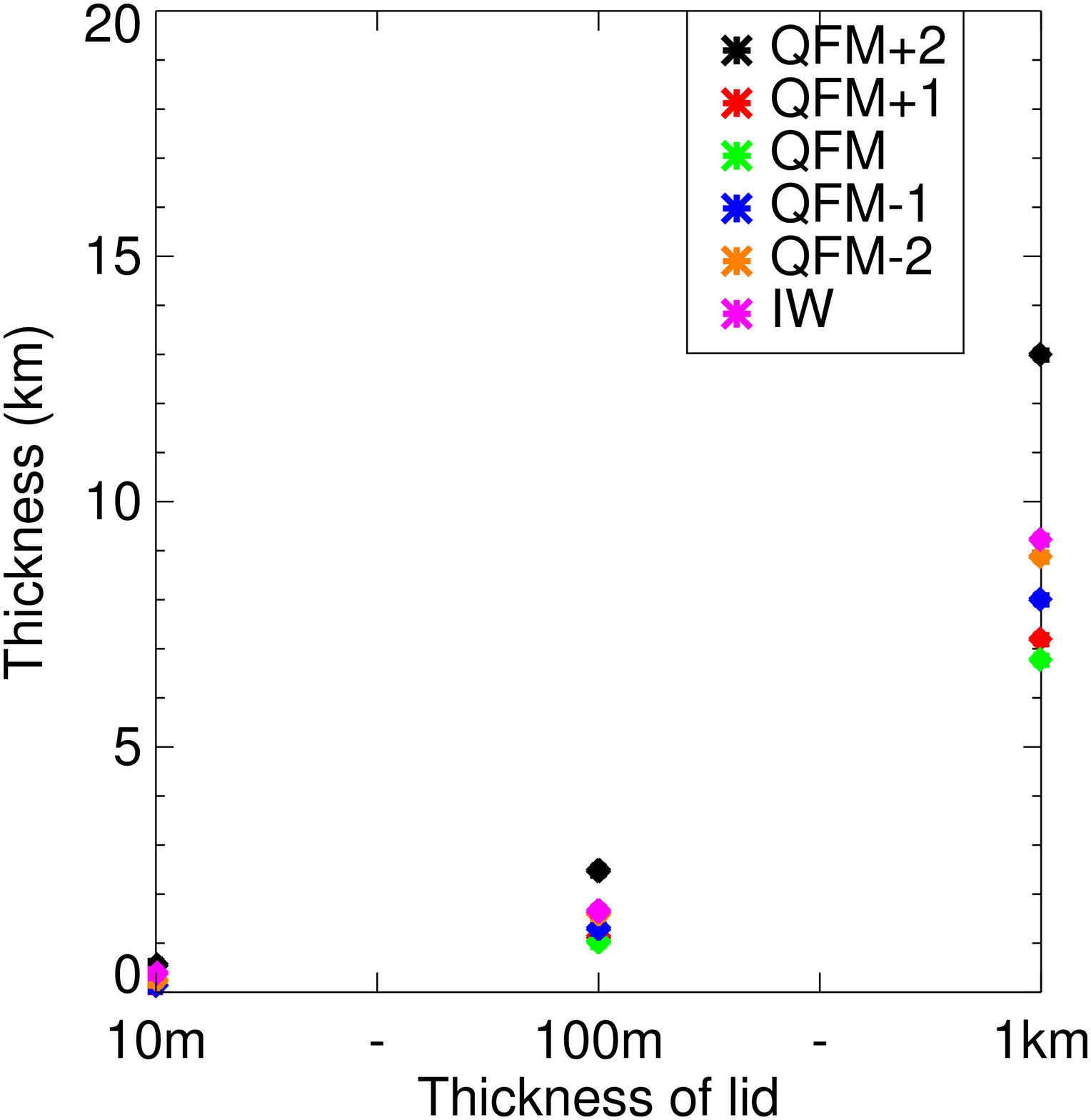}}
 \subfigure[diameter $a$=0.01cm]{
  \includegraphics[width=0.45\columnwidth,clip,bb=0.000000 0.000000 850.009733 850.009733]{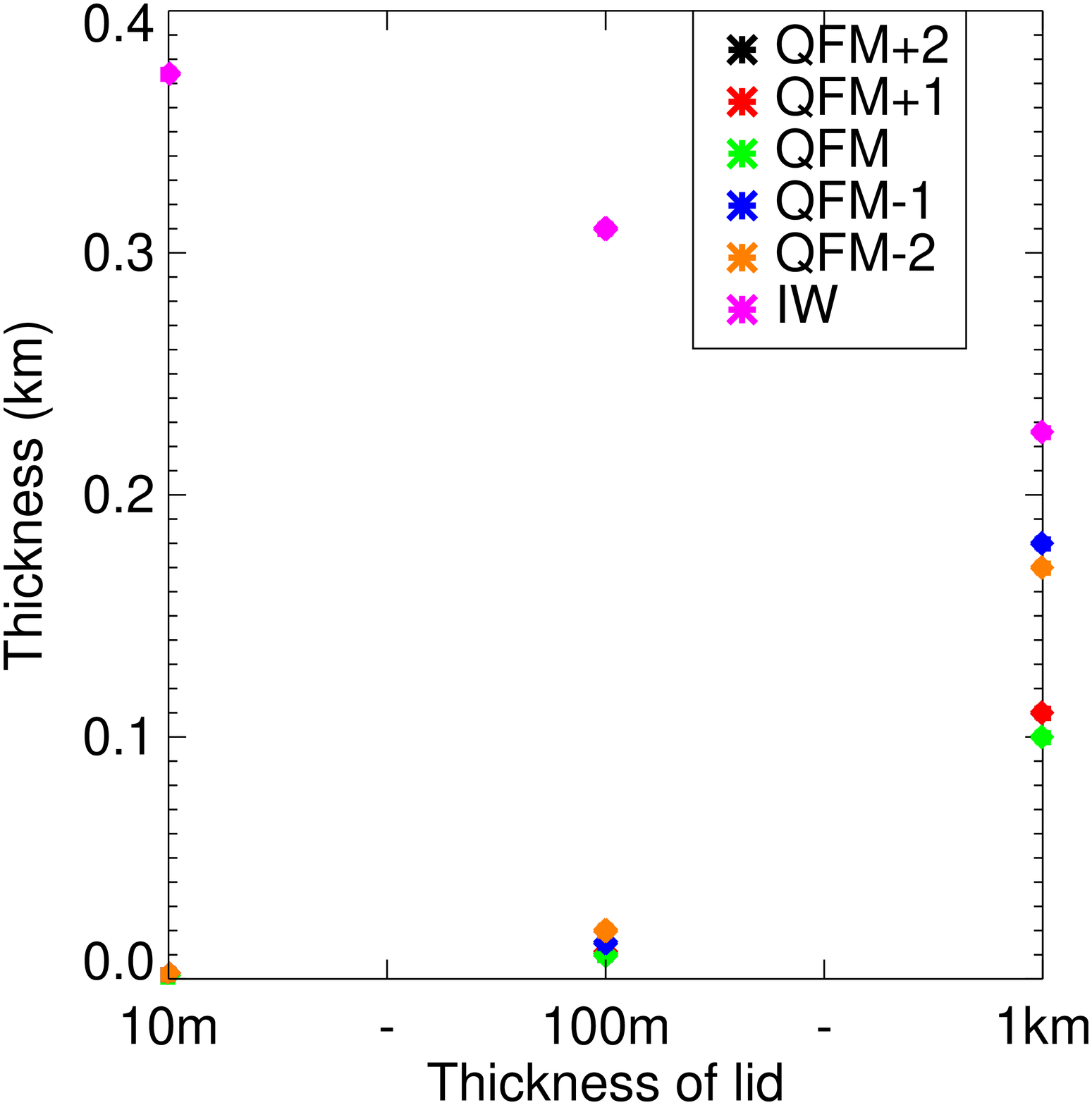}}
 \caption{Comparison of the thickness of the cumulate layer of the orthopyroxenite. The horizontal axis shows the thickness of the conductive lid. (a), (b), and (c) show the diameter of the crystal at 1 cm, 0.1 cm, and 0.01 cm respectively. The black, red, green, blue, orange, pink marks show the thickness of the orthopyroxene cumulate at ${\rm fO_2}=$ QFM $+$2, QFM $+$1, QFM, QFM $-$1, QFM $-2$, and IW, respectively.}
 \small{}
 \label{orthopyroxene}
\end{figure*}

In terms of the diogenite formation, the key point is the amount of orthopyroxene settled. It is important to note that orthopyroxene settles in the residual magma ocean in the early period of the crystallization. Our objective is to determine how the orthopyroxene crystallizes, settles, and forms a cumulate layer in the crystallization processes. Figure \ref{orthopyroxene} shows the thickness of the orthopyroxenite cumulate; the horizontal axis shows the thickness of the conductive lid. Figures \ref{orthopyroxene} a, b, and c are the cases for crystal diameters of 1 cm, 0.1 cm, and 0.01 cm, respectively. Colors represent $\rm{fO_{2}}$: black, QFM $+$ 2; red, QFM $+$ 1; green, QFM; blue, QFM $-$ 1; orange, QFM $-2$; and pink, IW. As shown in Fig. \ref{orthopyroxene}, small grain sizes of 0.01 cm and 0.1 cm and thin conductive lids of 10 m and 100 m create thin orthopyroxene cumulate layers of $\sim$1 km. If the conductive lid is thick, at 1 km, a thick orthopyroxene cumulate layer of $\sim$10 km is formed even though the crystal diameter is 0.1 cm. On the contrary, large crystals form thick cumulate layers of $\sim$20 km regardless of the thickness of the conductive lid. The fO$_{2}$ dependence of the cumulate thickness is shown in in Fig. \ref{orthopyroxene}b. In the figure, the orthopyroxene cumulate layer is thickest when the thickness of the lid is 1 km and fO$_{2}$ is at QFM +2 and is thinnest with the same lid thickness but with lower fO$_{2}$ (QFM). The difference in the two cases is only a factor of two. In Fig. \ref{orthopyroxene}c, the magma ocean with IW forms the maximum thickness of the orthopyroxenite regardless of the thickness of the lid. This is attributed to the lower solidus of the orthopyroxene with IW. Although such small differences exist, fO$_{2}$ does not have a crucial influence on the formation of the orthopyroxene cumulates.

As shown above, crystal size and thickness of the conductive lid critically influence the evolution of the magma ocean. The presence of diogenites (orthopyroxenites) requires the formation of an orthopyroxene cumulate layer with considerable thickness in Vesta, which was later broken up to be delivered to the Earth. Our simulation suggests that small ($\sim$0.01 cm) crystals cannot form a layer of orthopyroxene-dominated cumulate owing to the difficulty in settling in a turbulent magma ocean regardless of the lid thickness and fO$_{2}$. 

Although a thick cumulate layer of $\sim$20 km is formed with crystal diameters of 1 cm, the assumption of a uniform crystal size in the magma ocean would not be feasible. It is argued that crystals in a magma ocean are likely to be between 0.01 cm and 1 cm in diameter  \citep{1993LPI....24.1329S}. In such a case, 1 cm would be the maximum size. Although 1 cm crystals form a thick cumulate layer in a short duration, those 0.1 cm in size can form a cumulate layer $\sim$10 km thick when the conductive lid is as thick as 1 km. The efficiency of the settling will increase if mechanisms are present that lower the cooling speed, such as thick atmosphere formation caused by impact. 

\begin{figure*}
 \centering
\includegraphics[width=\columnwidth,clip,bb=0 0 1417 850]{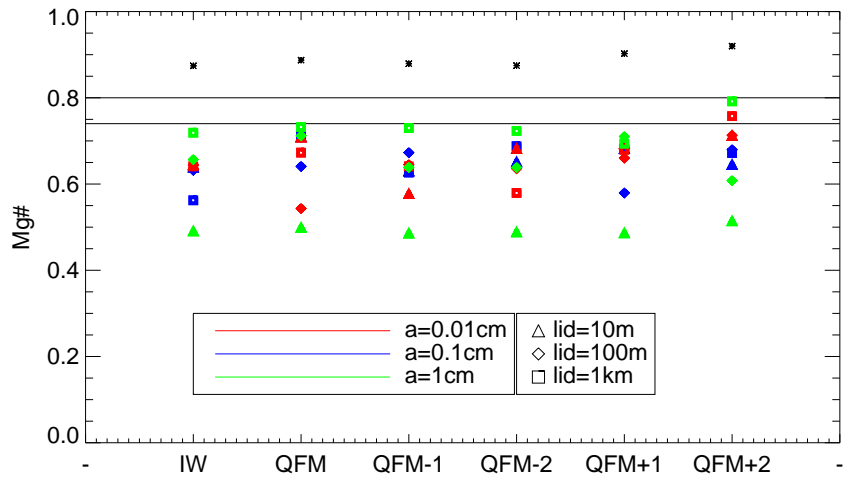}
 \caption{Maximum and minimum values of Mg$\#$ of orthopyroxene crystallized in the magma ocean with all parameters. The black star shows the maximum Mg$\#$ (the initial Mg$\#$ of this study). The colored symbols show the minimum value of Mg$\#$. The red, blue, and green symbols represent the grain sizes of 0.01 cm, 0.1 cm, and 1 cm, respectively. The thickness of the conductive lid is shown by the shape of the symbols; triangle, diamond, and square shapes indicate 10 m, 100 m, and 1 km, respectively. The black solid line shows the range of Mg$\#$ of the orthopyroxene following \cite{2000M&PS...35..901M}}.
 \small{}
 \label{mgfe}
\end{figure*}

The chemical composition of orthopyroxene in the cumulate layer is another crucial factor. Figure \ref{mgfe} shows the maximum and minimum values of the Mg$\#$ of orthopyroxene crystallized in the magma ocean considering all of the parameters. The black star shows the maximum Mg$\#$ (the initial Mg$\#$ of this study). The colored symbols show the minimum value of Mg$\#$. The red, blue, and green symbols represent grain sizes of 0.01 cm, 0.1 cm, and 1 cm, respectively. The thickness of the conductive lid is shown by the shape of the symbols; triangle, diamond, and square shapes represent 10 m, 100 m, and 1 km, respectively. The minimum Mg$\#$ of the orthopyroxene varied from 0.70 to 0.50 according to the parameters. Although fO$_{2}$ had little influence on the Mg$\#$, the thickness of the lid and the grain size affect the Mg$\#$ greatly. A thinner lid of 10 m with larger particles of 1 cm will produce a low Mg$\#$ because a thinner lid and larger particles accelerate the settling of the particles. As a result, Mg is preferentially incorporated into orthopyroxene in the magma ocean even after the crystallization of olivine, and the Mg$\#$ quickly decreases. Little difference was noted in the compositional evolution of orthopyroxene for the cases of 0.1 cm and 0.01 cm, which is also consistent with the fact that the settling and cumulate layer thickening were almost the same for the two grain sizes. It should be noted that this is the case only when the conductive lid is thin. If the lid is thicker than 1 km, a considerable difference is exhibited. Previous research \citep{2000M&PS...35..901M} has shown that the Mg$\#$ of diogenites varies between 0.74 and 0.80 (black solid line in Fig. \ref{mgfe}). Our calculation results of Mg$\#$ with large particles (1 cm) and a thicker lid (1 km) are almost consistent with the previous results. However, Fig. \ref{mgfe} indicates that oxidizing conditions such as ${\rm fO_{2}}$ at QFM +2 or QFM +1 cannot produce low Mg$\#$ orthopyroxene with small grain sizes. This is also consistent with the results of previous studies that show the redox state of Vesta to be reducing \citep{1996Icar..124..513R, 2013E&PSL.373...75P}.

\subsection{Comparison to the previous studies}  
\cite{1997M&PS...32..929R} suggested a model with 80$\%$ equilibrium crystallization followed by fractional crystallization. Their model produced a harzburgite mantle, cumulate eucrites, and noncumulate eucrites. Although their model produced orthopyroxenes, it also contains a certain amount of olivine ($>$50 wt$\%$). This is inconsistent with the composition of the diogenite that contains little olivine.

\cite{2013M&PS...48.2333M} also proposed a two-step model for differentiation of Vesta with 60--70$\%$ equilibrium crystallization followed by fractional crystallization of the residual melt. They showed that Vesta's mantle is composed of harzburgite and that the thickness of the crust is 30—41 km. Their model produced diogenite orthopyroxene, which, however, includes 10--20$\%$ olivine. Pure orthopyroxenite is difficult to produce.

\cite{2012espc.conf..909N, 2014E&PSL.395..267N} showed the importance of the presence of $^{26}$Al as a source of internal heating in the small parent body of HED meteorites, which was transported to the upper level once the mantle was partially melted to change the interior thermal structure. They used a spherically symmetric one-dimensional model that considered accretion, compaction, and melting. A shallow magma ocean with one to a few tens of kilometers was formed, and its lifetime was 10$^4$--10$^6$ years. They claimed that cumulate eucrites and diogenites might form through crystallization of the shallow magma ocean. Although they did not consider the chemical composition, our results support their shallow magma ocean model.

\section{Conclusion}
Our model has an advantage that fluid dynamics, thermal history, and chemical evolution of the Vesta magma ocean are consistently solved. Moreover, we showed that orthopyroxenite with a chemical composition consistent with diogenites is successfully reproduced through differentiation of the shallow magma ocean at a later stage with a limited range of cumulate crystal grain sizes and the thickness of the lid. We determined an appropriate condition of the magma ocean of Vesta for producing a thick orthopyroxenite layer by varying the size of the particles, the thickness of the conductive lid, and the ${\rm fO_{2}}$ as parameters. A plausible grain size is 1 cm, which produces a cumulate layer as thick as 10—20 km. The thickness of the conductive lid is another critical constraint. If the thickness of the lid is 1 km an orthopyroxenite layer of $\sim$20 km is produced when the particle size is 1 cm. The redox state is also important for diogenite formation. Although it has little influence on the thickness of the cumulate layer, it should be fairly reducing; ${\rm fO_{2}}$ at QFM or lower than two orders of magnitude would be desirable, which is consistent with observation and measurements. Our model concludes that the grain size of settled orthopyroxene was as large as 1 cm and that the lid was as thick as 1 km in the late stage of the magma ocean of Vesta.
\section{Acknowledgements}
This work was supported by a Grant-in-Aid for Scientific Research 25108003 (HN). The comments by an anonymous reviewer greatly improved the manuscript and are appreciated.
\section*{References}

%\bibliography{mybibfile}

\begin{thebibliography}{32}
\expandafter\ifx\csname natexlab\endcsname\relax\def\natexlab#1{#1}\fi
\providecommand{\url}[1]{\texttt{#1}}
\providecommand{\href}[2]{#2}
\providecommand{\path}[1]{#1}
\providecommand{\DOIprefix}{doi:}
\providecommand{\ArXivprefix}{arXiv:}
\providecommand{\URLprefix}{URL: }
\providecommand{\Pubmedprefix}{pmid:}
\providecommand{\doi}[1]{\href{http://dx.doi.org/#1}{\path{#1}}}
\providecommand{\Pubmed}[1]{\href{pmid:#1}{\path{#1}}}
\providecommand{\bibinfo}[2]{#2}
\ifx\xfnm\relax \def\xfnm[#1]{\unskip,\space#1}\fi
%Type = Inproceedings
\bibitem[{{Abe} and {Matsui}(1985)}]{1985LPSC...15..545A}
\bibinfo{author}{{Abe}, Y.}, \bibinfo{author}{{Matsui}, T.},
  \bibinfo{year}{1985}.
\newblock \bibinfo{title}{{The formation of an impact-generated H2O atmosphere
  and its implications for the early thermal history of the earth}}, in:
  \bibinfo{editor}{{Ryder}, G.}, \bibinfo{editor}{{Schubert}, G.} (Eds.),
  \bibinfo{booktitle}{Lunar and Planetary Science Conference Proceedings}, p.
  \bibinfo{pages}{545}.
%Type = Article
\bibitem[{Asimow and Ghiorso(1998)}]{asimow1998algorithmic}
\bibinfo{author}{Asimow, P.D.}, \bibinfo{author}{Ghiorso, M.S.},
  \bibinfo{year}{1998}.
\newblock \bibinfo{title}{Algorithmic modifications extending melts to
  calculate subsolidus phase relations}.
\newblock \bibinfo{journal}{American Mineralogist} \bibinfo{volume}{83},
  \bibinfo{pages}{1127--1132}.
%Type = Article
\bibitem[{{Boss}(1986)}]{1986Sci...231..341B}
\bibinfo{author}{{Boss}, A.P.}, \bibinfo{year}{1986}.
\newblock \bibinfo{title}{{The origin of the moon}}.
\newblock \bibinfo{journal}{Science} \bibinfo{volume}{231},
  \bibinfo{pages}{341--345}.
\newblock \DOIprefix\doi{10.1126/science.231.4736.341}.
%Type = Article
\bibitem[{{De Sanctis} et~al.(2013){De Sanctis}, {Ammannito}, {Capria},
  {Capaccioni}, {Combe}, {Frigeri}, {Longobardo}, {Magni}, {Marchi}, {McCord},
  {Palomba}, {Tosi}, {Zambon}, {Carraro}, {Fonte}, {Li}, {McFadden},
  {Mittlefehldt}, {Pieters}, {Jaumann}, {Stephan}, {Raymond} and
  {Russell}}]{2013M&PS...48.2166D}
\bibinfo{author}{{De Sanctis}, M.C.}, \bibinfo{author}{{Ammannito}, E.},
  \bibinfo{author}{{Capria}, M.T.}, \bibinfo{author}{{Capaccioni}, F.},
  \bibinfo{author}{{Combe}, J.P.}, \bibinfo{author}{{Frigeri}, A.},
  \bibinfo{author}{{Longobardo}, A.}, \bibinfo{author}{{Magni}, G.},
  \bibinfo{author}{{Marchi}, S.}, \bibinfo{author}{{McCord}, T.B.},
  \bibinfo{author}{{Palomba}, E.}, \bibinfo{author}{{Tosi}, F.},
  \bibinfo{author}{{Zambon}, F.}, \bibinfo{author}{{Carraro}, F.},
  \bibinfo{author}{{Fonte}, S.}, \bibinfo{author}{{Li}, Y.J.},
  \bibinfo{author}{{McFadden}, L.A.}, \bibinfo{author}{{Mittlefehldt}, D.W.},
  \bibinfo{author}{{Pieters}, C.M.}, \bibinfo{author}{{Jaumann}, R.},
  \bibinfo{author}{{Stephan}, K.}, \bibinfo{author}{{Raymond}, C.A.},
  \bibinfo{author}{{Russell}, C.T.}, \bibinfo{year}{2013}.
\newblock \bibinfo{title}{{Vesta's mineralogical composition as revealed by the
  visible and infrared spectrometer on Dawn}}.
\newblock \bibinfo{journal}{Meteoritics and Planetary Science}
  \bibinfo{volume}{48}, \bibinfo{pages}{2166--2184}.
\newblock \DOIprefix\doi{10.1111/maps.12138}.
%Type = Article
\bibitem[{{De Sanctis} et~al.(2011){De Sanctis}, {Coradini}, {Ammannito},
  {Filacchione}, {Capria}, {Fonte}, {Magni}, {Barbis}, {Bini}, {Dami},
  {Ficai-Veltroni} and {Preti}}]{2011SSRv..163..329D}
\bibinfo{author}{{De Sanctis}, M.C.}, \bibinfo{author}{{Coradini}, A.},
  \bibinfo{author}{{Ammannito}, E.}, \bibinfo{author}{{Filacchione}, G.},
  \bibinfo{author}{{Capria}, M.T.}, \bibinfo{author}{{Fonte}, S.},
  \bibinfo{author}{{Magni}, G.}, \bibinfo{author}{{Barbis}, A.},
  \bibinfo{author}{{Bini}, A.}, \bibinfo{author}{{Dami}, M.},
  \bibinfo{author}{{Ficai-Veltroni}, I.}, \bibinfo{author}{{Preti}, G.},
  \bibinfo{year}{2011}.
\newblock \bibinfo{title}{{The VIR Spectrometer}}.
\newblock \bibinfo{journal}{\ssr} \bibinfo{volume}{163},
  \bibinfo{pages}{329--369}.
\newblock \DOIprefix\doi{10.1007/s11214-010-9668-5}.
%Type = Article
\bibitem[{{Elkins-Tanton}(2008)}]{2008E&PSL.271..181E}
\bibinfo{author}{{Elkins-Tanton}, L.T.}, \bibinfo{year}{2008}.
\newblock \bibinfo{title}{{Linked magma ocean solidification and atmospheric
  growth for Earth and Mars}}.
\newblock \bibinfo{journal}{Earth and Planetary Science Letters}
  \bibinfo{volume}{271}, \bibinfo{pages}{181--191}.
\newblock \DOIprefix\doi{10.1016/j.epsl.2008.03.062}.
%Type = Article
\bibitem[{{Elkins-Tanton}(2012)}]{2012AREPS..40..113E}
\bibinfo{author}{{Elkins-Tanton}, L.T.}, \bibinfo{year}{2012}.
\newblock \bibinfo{title}{{Magma Oceans in the Inner Solar System}}.
\newblock \bibinfo{journal}{Annual Review of Earth and Planetary Sciences}
  \bibinfo{volume}{40}, \bibinfo{pages}{113--139}.
\newblock \DOIprefix\doi{10.1146/annurev-earth-042711-105503}.
%Type = Article
\bibitem[{Ghiorso and Sack(1995)}]{ghiorso1995chemical}
\bibinfo{author}{Ghiorso, M.S.}, \bibinfo{author}{Sack, R.O.},
  \bibinfo{year}{1995}.
\newblock \bibinfo{title}{Chemical mass transfer in magmatic processes iv. a
  revised and internally consistent thermodynamic model for the interpolation
  and extrapolation of liquid-solid equilibria in magmatic systems at elevated
  temperatures and pressures}.
\newblock \bibinfo{journal}{Contributions to Mineralogy and Petrology}
  \bibinfo{volume}{119}, \bibinfo{pages}{197--212}.
%Type = Article
\bibitem[{{Grossmann} and {Lohse}(2000)}]{so24778}
\bibinfo{author}{{Grossmann}, S.}, \bibinfo{author}{{Lohse}, D.},
  \bibinfo{year}{2000}.
\newblock \bibinfo{title}{Scaling in thermal convection: A unifying view}.
\newblock \bibinfo{journal}{Journal of Fluid Mechanics} \bibinfo{volume}{407},
  \bibinfo{pages}{27--56}.
\newblock \URLprefix \url{http://doc.utwente.nl/24778/}.
%Type = Article
\bibitem[{{Kraichnan}(1962)}]{1962PhFl....5.1374K}
\bibinfo{author}{{Kraichnan}, R.H.}, \bibinfo{year}{1962}.
\newblock \bibinfo{title}{{Turbulent Thermal Convection at Arbitrary Prandtl
  Number}}.
\newblock \bibinfo{journal}{Physics of Fluids} \bibinfo{volume}{5},
  \bibinfo{pages}{1374--1389}.
\newblock \DOIprefix\doi{10.1063/1.1706533}.
%Type = Article
\bibitem[{Lejeune and Richet(1995)}]{lejeune1995rheology}
\bibinfo{author}{Lejeune, A.M.}, \bibinfo{author}{Richet, P.},
  \bibinfo{year}{1995}.
\newblock \bibinfo{title}{Rheology of crystal-bearing silicate melts: An
  experimental study at high viscosities}.
\newblock \bibinfo{journal}{Journal of Geophysical Research: Solid Earth}
  \bibinfo{volume}{100}, \bibinfo{pages}{4215--4229}.
%Type = Article
\bibitem[{{Mandler} and {Elkins-Tanton}(2013)}]{2013M&PS...48.2333M}
\bibinfo{author}{{Mandler}, B.E.}, \bibinfo{author}{{Elkins-Tanton}, L.T.},
  \bibinfo{year}{2013}.
\newblock \bibinfo{title}{{The origin of eucrites, diogenites, and olivine
  diogenites: Magma ocean crystallization and shallow magma chamber processes
  on Vesta}}.
\newblock \bibinfo{journal}{Meteoritics and Planetary Science}
  \bibinfo{volume}{48}, \bibinfo{pages}{2333--2349}.
\newblock \DOIprefix\doi{10.1111/maps.12135}.
%Type = Article
\bibitem[{Martin and Nokes(1989)}]{martin1989fluid}
\bibinfo{author}{Martin, D.}, \bibinfo{author}{Nokes, R.},
  \bibinfo{year}{1989}.
\newblock \bibinfo{title}{A fluid-dynamical study of crystal settling in
  convecting magmas}.
\newblock \bibinfo{journal}{Journal of Petrology} \bibinfo{volume}{30},
  \bibinfo{pages}{1471--1500}.
%Type = Article
\bibitem[{{Mittlefehldt}(2000)}]{2000M&PS...35..901M}
\bibinfo{author}{{Mittlefehldt}, D.W.}, \bibinfo{year}{2000}.
\newblock \bibinfo{title}{{Petrology and geochemistry of the Elephant Moraine
  A79002 diogenite: A genomict breccia containing a magnesian harzburgite
  component}}.
\newblock \bibinfo{journal}{Meteoritics and Planetary Science}
  \bibinfo{volume}{35}, \bibinfo{pages}{901--912}.
\newblock \DOIprefix\doi{10.1111/j.1945-5100.2000.tb01479.x}.
%Type = Article
\bibitem[{Van~der Molen and Paterson(1979)}]{van1979experimental}
\bibinfo{author}{Van~der Molen, I.}, \bibinfo{author}{Paterson, M.},
  \bibinfo{year}{1979}.
\newblock \bibinfo{title}{Experimental deformation of partially-melted
  granite}.
\newblock \bibinfo{journal}{Contributions to Mineralogy and Petrology}
  \bibinfo{volume}{70}, \bibinfo{pages}{299--318}.
%Type = Article
\bibitem[{{Moskovitz} and {Gaidos}(2011)}]{2011M&PS...46..903M}
\bibinfo{author}{{Moskovitz}, N.}, \bibinfo{author}{{Gaidos}, E.},
  \bibinfo{year}{2011}.
\newblock \bibinfo{title}{{Differentiation of planetesimals and the thermal
  consequences of melt migration}}.
\newblock \bibinfo{journal}{Meteoritics and Planetary Science}
  \bibinfo{volume}{46}, \bibinfo{pages}{903--918}.
\newblock \DOIprefix\doi{10.1111/j.1945-5100.2011.01201.x},
  \href{http://arxiv.org/abs/1101.4165}{\tt arXiv:1101.4165}.
%Type = Inproceedings
\bibitem[{{Neumann} et~al.(2012){Neumann}, {Breuer} and
  {Spohn}}]{2012espc.conf..909N}
\bibinfo{author}{{Neumann}, W.}, \bibinfo{author}{{Breuer}, D.},
  \bibinfo{author}{{Spohn}, T.}, \bibinfo{year}{2012}.
\newblock \bibinfo{title}{{Numerical modelling of accretion, sintering and
  differentiation of asteroid 4 Vesta}}, in: \bibinfo{booktitle}{European
  Planetary Science Congress 2012}, p. \bibinfo{pages}{909}.
%Type = Article
\bibitem[{{Neumann} et~al.(2014){Neumann}, {Breuer} and
  {Spohn}}]{2014E&PSL.395..267N}
\bibinfo{author}{{Neumann}, W.}, \bibinfo{author}{{Breuer}, D.},
  \bibinfo{author}{{Spohn}, T.}, \bibinfo{year}{2014}.
\newblock \bibinfo{title}{{Differentiation of Vesta: Implications for a shallow
  magma ocean}}.
\newblock \bibinfo{journal}{Earth and Planetary Science Letters}
  \bibinfo{volume}{395}, \bibinfo{pages}{267--280}.
\newblock \DOIprefix\doi{10.1016/j.epsl.2014.03.033},
  \href{http://arxiv.org/abs/1402.3103}{\tt arXiv:1402.3103}.
%Type = Article
\bibitem[{{Prettyman} et~al.(2012){Prettyman}, {Mittlefehldt}, {Yamashita},
  {Lawrence}, {Beck}, {Feldman}, {McCoy}, {McSween}, {Toplis}, {Titus},
  {Tricarico}, {Reedy}, {Hendricks}, {Forni}, {Le Corre}, {Li}, {Mizzon},
  {Reddy}, {Raymond} and {Russell}}]{2012Sci...338..242P}
\bibinfo{author}{{Prettyman}, T.H.}, \bibinfo{author}{{Mittlefehldt}, D.W.},
  \bibinfo{author}{{Yamashita}, N.}, \bibinfo{author}{{Lawrence}, D.J.},
  \bibinfo{author}{{Beck}, A.W.}, \bibinfo{author}{{Feldman}, W.C.},
  \bibinfo{author}{{McCoy}, T.J.}, \bibinfo{author}{{McSween}, H.Y.},
  \bibinfo{author}{{Toplis}, M.J.}, \bibinfo{author}{{Titus}, T.N.},
  \bibinfo{author}{{Tricarico}, P.}, \bibinfo{author}{{Reedy}, R.C.},
  \bibinfo{author}{{Hendricks}, J.S.}, \bibinfo{author}{{Forni}, O.},
  \bibinfo{author}{{Le Corre}, L.}, \bibinfo{author}{{Li}, J.Y.},
  \bibinfo{author}{{Mizzon}, H.}, \bibinfo{author}{{Reddy}, V.},
  \bibinfo{author}{{Raymond}, C.A.}, \bibinfo{author}{{Russell}, C.T.},
  \bibinfo{year}{2012}.
\newblock \bibinfo{title}{{Elemental Mapping by Dawn Reveals Exogenic H in
  Vesta's Regolith}}.
\newblock \bibinfo{journal}{Science} \bibinfo{volume}{338},
  \bibinfo{pages}{242--}.
\newblock \DOIprefix\doi{10.1126/science.1225354}.
%Type = Article
\bibitem[{{Pringle} et~al.(2013){Pringle}, {Savage}, {Badro}, {Barrat} and
  {Moynier}}]{2013E&PSL.373...75P}
\bibinfo{author}{{Pringle}, E.A.}, \bibinfo{author}{{Savage}, P.S.},
  \bibinfo{author}{{Badro}, J.}, \bibinfo{author}{{Barrat}, J.A.},
  \bibinfo{author}{{Moynier}, F.}, \bibinfo{year}{2013}.
\newblock \bibinfo{title}{{Redox state during core formation on asteroid
  4-Vesta}}.
\newblock \bibinfo{journal}{Earth and Planetary Science Letters}
  \bibinfo{volume}{373}, \bibinfo{pages}{75--82}.
\newblock \DOIprefix\doi{10.1016/j.epsl.2013.04.012}.
%Type = Article
\bibitem[{{Righter} and {Drake}(1996)}]{1996Icar..124..513R}
\bibinfo{author}{{Righter}, K.}, \bibinfo{author}{{Drake}, M.J.},
  \bibinfo{year}{1996}.
\newblock \bibinfo{title}{{Core Formation in Earth's Moon, Mars, and Vesta}}.
\newblock \bibinfo{journal}{\icarus} \bibinfo{volume}{124},
  \bibinfo{pages}{513--529}.
\newblock \DOIprefix\doi{10.1006/icar.1996.0227}.
%Type = Article
\bibitem[{{Righter} and {Drake}(1997)}]{1997M&PS...32..929R}
\bibinfo{author}{{Righter}, K.}, \bibinfo{author}{{Drake}, M.J.},
  \bibinfo{year}{1997}.
\newblock \bibinfo{title}{{A magma ocean on Vesta: Core formation and
  petrogenesis of eucrites and diogenites}}.
\newblock \bibinfo{journal}{Meteoritics and Planetary Science}
  \bibinfo{volume}{32}, \bibinfo{pages}{929--944}.
\newblock \DOIprefix\doi{10.1111/j.1945-5100.1997.tb01582.x}.
%Type = Article
\bibitem[{{Russell} et~al.(2012){Russell}, {Raymond}, {Coradini}, {McSween},
  {Zuber}, {Nathues}, {De Sanctis}, {Jaumann}, {Konopliv}, {Preusker}, {Asmar},
  {Park}, {Gaskell}, {Keller}, {Mottola}, {Roatsch}, {Scully}, {Smith},
  {Tricarico}, {Toplis}, {Christensen}, {Feldman}, {Lawrence}, {McCoy},
  {Prettyman}, {Reedy}, {Sykes} and {Titus}}]{2012Sci...336..684R}
\bibinfo{author}{{Russell}, C.T.}, \bibinfo{author}{{Raymond}, C.A.},
  \bibinfo{author}{{Coradini}, A.}, \bibinfo{author}{{McSween}, H.Y.},
  \bibinfo{author}{{Zuber}, M.T.}, \bibinfo{author}{{Nathues}, A.},
  \bibinfo{author}{{De Sanctis}, M.C.}, \bibinfo{author}{{Jaumann}, R.},
  \bibinfo{author}{{Konopliv}, A.S.}, \bibinfo{author}{{Preusker}, F.},
  \bibinfo{author}{{Asmar}, S.W.}, \bibinfo{author}{{Park}, R.S.},
  \bibinfo{author}{{Gaskell}, R.}, \bibinfo{author}{{Keller}, H.U.},
  \bibinfo{author}{{Mottola}, S.}, \bibinfo{author}{{Roatsch}, T.},
  \bibinfo{author}{{Scully}, J.E.C.}, \bibinfo{author}{{Smith}, D.E.},
  \bibinfo{author}{{Tricarico}, P.}, \bibinfo{author}{{Toplis}, M.J.},
  \bibinfo{author}{{Christensen}, U.R.}, \bibinfo{author}{{Feldman}, W.C.},
  \bibinfo{author}{{Lawrence}, D.J.}, \bibinfo{author}{{McCoy}, T.J.},
  \bibinfo{author}{{Prettyman}, T.H.}, \bibinfo{author}{{Reedy}, R.C.},
  \bibinfo{author}{{Sykes}, M.E.}, \bibinfo{author}{{Titus}, T.N.},
  \bibinfo{year}{2012}.
\newblock \bibinfo{title}{{Dawn at Vesta: Testing the Protoplanetary
  Paradigm}}.
\newblock \bibinfo{journal}{Science} \bibinfo{volume}{336},
  \bibinfo{pages}{684--}.
\newblock \DOIprefix\doi{10.1126/science.1219381}.
%Type = Article
\bibitem[{{Siggia}(1994)}]{1994AnRFM..26..137S}
\bibinfo{author}{{Siggia}, E.D.}, \bibinfo{year}{1994}.
\newblock \bibinfo{title}{{High rayleigh number convection}}.
\newblock \bibinfo{journal}{Annual Review of Fluid Mechanics}
  \bibinfo{volume}{26}, \bibinfo{pages}{137--168}.
\newblock \DOIprefix\doi{10.1146/annurev.fl.26.010194.001033}.
%Type = Inbook
\bibitem[{{Solomatov}(2000)}]{2000orem.book..323S}
\bibinfo{author}{{Solomatov}, V.S.}, \bibinfo{year}{2000}.
\newblock \bibinfo{title}{{Fluid Dynamics of a Terrestrial Magma Ocean}}.
\newblock pp. \bibinfo{pages}{323--338}.
%Type = Article
\bibitem[{Solomatov(2007)}]{solomatov2007magma}
\bibinfo{author}{Solomatov, V.S.}, \bibinfo{year}{2007}.
\newblock \bibinfo{title}{Magma oceans and primordial mantle differentiation}.
\newblock \bibinfo{journal}{Treatise on Geophysics} \bibinfo{volume}{9},
  \bibinfo{pages}{91--120}.
%Type = Inproceedings
\bibitem[{{Solomatov} and {Stevenson}(1993a)}]{1993LPI....24.1329S}
\bibinfo{author}{{Solomatov}, V.S.}, \bibinfo{author}{{Stevenson}, D.J.},
  \bibinfo{year}{1993}a.
\newblock \bibinfo{title}{{Differentiation of magma oceans and the thickness of
  the depleted layer on Venus}}, in: \bibinfo{booktitle}{Lunar and Planetary
  Science Conference}, pp. \bibinfo{pages}{1329--1330}.
%Type = Article
\bibitem[{{Solomatov} and {Stevenson}(1993b)}]{1993JGR....98.5407S}
\bibinfo{author}{{Solomatov}, V.S.}, \bibinfo{author}{{Stevenson}, D.J.},
  \bibinfo{year}{1993}b.
\newblock \bibinfo{title}{{Kinetics of crystal growth in a terrestrial magma
  ocean}}.
\newblock \bibinfo{journal}{\jgr} \bibinfo{volume}{98},
  \bibinfo{pages}{5407--5418}.
\newblock \DOIprefix\doi{10.1029/92JE02839}.
%Type = Article
\bibitem[{{Solomatov} and {Stevenson}(1993c)}]{1993JGR....98.5375S}
\bibinfo{author}{{Solomatov}, V.S.}, \bibinfo{author}{{Stevenson}, D.J.},
  \bibinfo{year}{1993}c.
\newblock \bibinfo{title}{{Suspension in convective layers and style of
  differentiation of a terrestrial magma ocean}}.
\newblock \bibinfo{journal}{\jgr} \bibinfo{volume}{98},
  \bibinfo{pages}{5375--5390}.
\newblock \DOIprefix\doi{10.1029/92JE02948}.
%Type = Article
\bibitem[{{Stevenson}(1987)}]{1987AREPS..15..271S}
\bibinfo{author}{{Stevenson}, D.J.}, \bibinfo{year}{1987}.
\newblock \bibinfo{title}{{Origin of the moon - The collision hypothesis}}.
\newblock \bibinfo{journal}{Annual Review of Earth and Planetary Sciences}
  \bibinfo{volume}{15}, \bibinfo{pages}{271--315}.
\newblock \DOIprefix\doi{10.1146/annurev.ea.15.050187.001415}.
%Type = Article
\bibitem[{{Thomas} et~al.(1997){Thomas}, {Binzel}, {Gaffey}, {Storrs}, {Wells}
  and {Zellner}}]{1997Sci...277.1492T}
\bibinfo{author}{{Thomas}, P.C.}, \bibinfo{author}{{Binzel}, R.P.},
  \bibinfo{author}{{Gaffey}, M.J.}, \bibinfo{author}{{Storrs}, A.D.},
  \bibinfo{author}{{Wells}, E.N.}, \bibinfo{author}{{Zellner}, B.H.},
  \bibinfo{year}{1997}.
\newblock \bibinfo{title}{{Impact excavation on asteroid 4 Vesta: Hubble Space
  Telescope results}}.
\newblock \bibinfo{journal}{Science} \bibinfo{volume}{277},
  \bibinfo{pages}{1492--1495}.
\newblock \DOIprefix\doi{10.1126/science.277.5331.1492}.
%Type = Article
\bibitem[{{Zuber} et~al.(2011){Zuber}, {McSween}, {Binzel}, {Elkins-Tanton},
  {Konopliv}, {Pieters} and {Smith}}]{2011SSRv..163...77Z}
\bibinfo{author}{{Zuber}, M.T.}, \bibinfo{author}{{McSween}, H.Y.},
  \bibinfo{author}{{Binzel}, R.P.}, \bibinfo{author}{{Elkins-Tanton}, L.T.},
  \bibinfo{author}{{Konopliv}, A.S.}, \bibinfo{author}{{Pieters}, C.M.},
  \bibinfo{author}{{Smith}, D.E.}, \bibinfo{year}{2011}.
\newblock \bibinfo{title}{{Origin, Internal Structure and Evolution of 4
  Vesta}}.
\newblock \bibinfo{journal}{\ssr} \bibinfo{volume}{163},
  \bibinfo{pages}{77--93}.
\newblock \DOIprefix\doi{10.1007/s11214-011-9806-8}.

\end{thebibliography}

\clearpage
\begin{table*}
\centering
 \caption{Physical parameters used in the model}
  \begin{tabular}{lcccc} \hline
  Quantity & Symbol & Value & Unit&Reference\\ \hline  \hline
  Thermal expansion& $\alpha$ & $5.0 \times 10^{-5}$ & ${\rm K^{-1}}$ &(1)\\
 Thermal capacity &$c_{p}$& $1.0\times10^{3}$ & J ${\rm kg^{-1}K^{-1}}$&(1) \\
  Gravity & $g$ & 0.32 & ${\rm m\, s^{-2}}$&(2) \\
  Stefan-Boltzmann constant &$\sigma$& $5.67\times10^{-8}$ &${\rm J\, m^{-2}K^{-4}}$&--\\
  $^{26}{\rm Al}$ half-life &$\tau_{1/2}$&$7.17\times10^5$&year&(3) \\
  Latent heat &$Q_{{\rm latent}}$&$4.0\times10^{5}$&${\rm J\, kg^{-1}}$&(3)\\
   Initial magma ocean depth & $h$ &$ 5.0\times 10^{4}$& m &--\\
  Diameter of the crystal &$a$&$1.0\times 10^{-4},1.0\times 10^{-3}, 1.0\times 10^{-2}$&m&--\\
  Thickness of the lid &$l$&10,100,1000&m&-- \\ \hline
    \end{tabular}
     \label{tb:parameter}
    {\bf References.}  (1) \cite{2000orem.book..323S,solomatov2007magma}, (2) \cite{2011SSRv..163...77Z}, (3) \cite{2012espc.conf..909N,2014E&PSL.395..267N} 
       \end{table*}
\clearpage

\begin{table*}
\caption{Initial composition of the magma ocean}
\centering
\begin{tabular}{lc|cccccc}\hline
&Silicate portion& QFM+2 & QFM+1 & QFM & QFM-1 &QFM-2 &IW\\ \hline
${\rm SiO_{2}}$ & 49.7 & 51.2 & 53.2 &55.2 &55.8 &56.5 &56.9\\
${\rm TiO_{2}}$ & 0.13 & 0.24 & 0.25 & 0.26 & 0.26 & 0.26 &0.26\\
${\rm Al_{2}O_{3}}$ & 2.87 & 5.08 & 5.34 & 5.65 & 5.58 & 5.63&5.69 \\
${\rm Cr_{2}O_{3}}$ & 0.71 & 0.17 & 0.26 & 0.31 & 0.39 & 0.43 &0.45\\
FeO & 13.96 & 14.3 & 14.5 & 14.2 & 14.2 & 14.1&14.0\\
MnO & 0.44 & 0.48 & 0.44 & 0.41 & 0.40 & 0.39 &0.39\\
MgO & 30.31 & 16.38 & 15.6 & 14.6 & 14.8 & 14.6 &14.4\\
CaO & 2.28 & 4.27 & 4.32 & 4.45 & 4.39 & 4.41&4.45\\
${\rm Na_{2}O}$ & 1.26 & 2.48 & 2.50 & 2.58 & 2.53 & 2.54&2.56 \\
${\rm K_{2}O}$ & 0.13 & 0.26 & 0.26 & 0.27 & 0.26 & 0.26&0.26\\ \hline
\end{tabular}
\label{tb:composition}
\begin{flushleft}
Silicate portion shows the silicate composition of Fig.\ref{model}(a). This value comes from L chondrite after differentiation of the olivine mantle following \cite{1997M&PS...32..929R}.\\
QFM+2, QFM+1, QFM, QFM-1, QFM-2, and IW are the initial composition of the 50 km deep magma ocean of Fig. \ref{model}(b) after extraction of olivine crystallized at each redox state, which was calculated with MELTS.
\end{flushleft}
\end{table*}
\clearpage

\clearpage
\clearpage

 \clearpage

\end{document}